\documentclass[aps,prb,twocolumn,showpacs,superscriptaddress]{revtex4}%
\usepackage{graphicx}
\usepackage{dcolumn}
\usepackage{bm}
\usepackage{amssymb}
\usepackage{amsmath}
\usepackage{epstopdf}
\usepackage{epsfig}
\usepackage{amsfonts}%
\providecommand{\U}[1]{\protect\rule{.1in}{.1in}}
\begin{document}
\title{Site-wise manipulations and Mott insulator-superfluid transition of interacting photons using superconducting circuit simulators}
\author{Xiuhao Deng}
\affiliation{School of Natural Sciences, University of California Merced, Merced,
California 95343, USA}
\author{Chunjing Jia}
\affiliation{Department of Applied Physics, Stanford University, California 94305, USA}
\affiliation{Stanford Institute for Materials and Energy Sciences, SLAC National
Accelerator Laboratory, 2575 Sand Hill Road, Menlo Park, California 94025, USA}
\author{Chih-Chun Chien}
\email{cchien5@ucmerced.edu}
\affiliation{School of Natural Sciences, University of California Merced, Merced,
California 95343, USA}


\begin{abstract}
The Bose Hubbard model (BHM) of interacting bosons in a lattice has been a
paradigm in many-body physics, and it exhibits a Mott insulator (MI)-superfluid (SF) transition at integer filling. Here a quantum simulator of the BHM using a 
superconducting circuit is proposed. Specifically, a superconducting
transmission line resonator supporting microwave photons is coupled to a
charge qubit to form one site of the BHM, and adjacent sites are connected by
a tunable coupler. To obtain a mapping from the superconducting circuit to the
BHM, we focus on the dispersive regime where the excitations remain
photon-like. Standard perturbation theory is implemented to locate the
parameter range where the MI-SF transition may be simulated. This simulator allows
single-site manipulations and we illustrate this feature by considering two
scenarios where a single-site manipulation can drive a MI-SF transition. The transition can be analyzed by mean-field analyses, and the exact diagonalization was
implemented to provide accurate results. The variance of the photon density and the
fidelity metric clearly show signatures of the transition. Experimental
realizations and other possible applications of this simulator are also discussed.
\end{abstract}
\date{\today}
\maketitle



\section{Introduction}

Intense research has been focused on simulating complex matter using
well-controlled quantum systems in order to better understand their behavior
and create useful analogues \cite{QuantumSimulation,SimulationColdAtom,
SimulationTrapIon,SimulationDynamicTrapIon, SimulationDiamond,
SimulationPhotonic}. Successful examples include cold atoms trapped in optical
potentials \cite{SimulationColdAtom}, trapped ions
\cite{SimulationTrapIon,SimulationDynamicTrapIon}, spins in defects in
diamonds \cite{SimulationDiamond}, photonic arrays \cite{SimulationPhotonic},
etc. Recently, another class of quantum simulators based on superconducting
circuits opens more opportunities
\cite{DevoretReview2,SimulationWithSCCircuit,MooijSCNetwork,FazioQuanPhasTSCCircuit,BHMJJarrays}%
, which is made possible due to progresses in fabricating well-designed
circuits on chips. In those superconducting circuits, dissipation and
decoherence have been suppressed significantly \cite{DevoretReview2,Q factor
3D Schoelkopf}. Moreover, interacting superconducting qubits or resonators can
be fabricated on a chip, where quantum error-correction encoding and high
fidelity operations have been realized \cite{SurfaceCode Nature, SurfaceCode
NatureComm}. Various designs of couplers for connecting different qubits or
resonators with wide tuning ranges have also been demonstrated
\cite{GrossTunableExper,FastTunableCoupler,YuChenTunableCoupler}. Those
progresses in superconducting circuits provide a promising perspective of
scalable superconducting circuits as quantum simulators for many-body systems,
which may be bosonic\cite{SimulationWithSCCircuit,cQED lattice,Network
nonlinear SC TLR chain,MI1Dvortices} or fermionic\cite{Ising chain,Majorana SC
qubit} in nature.

The Bose-Hubbard Model (BHM) has been a paradigm in many-body theories, and
the Mott insulator-superfluid (MI-SF) phase transition associated with the BHM
has been of broad interest \cite{BHM,SimulationColdAtom}. This transition was
observed unambiguously in cold atoms trapped in optical lattices and can be
probed with single-atom resolutions \cite{GreinerNat, GreinerSci}. On the
other hand, a theoretical framework for obtaining the BHM using the
Jaynes-Cummings Hubbard Model has been established \cite{CQED MI SF TR,CQED
BHM JC array}. Simulating this general model in cavity arrays has been
proposed \cite{BHM Cavity Array,PhotonQPhaseTR,CQED MI SF
TR,HartmannPlenioPRL2007}. One may envision that introducing inhomogeneity
into the BHM parameters can lead to richer physics, some of which has been
explored in Refs.~\onlinecite{SLiang1D,Edgestate 1D Superlattice}. Simulating
those phenomena requires tunability of single-site parameters, which could be
hard in current available simulators
\cite{QuantumSimulation,SimulationColdAtom,SimulationTrapIon,SimulationDynamicTrapIon,SimulationDiamond}%
.

As a candidate of quantum simulators, superconducting circuit has the
following additional features
\cite{QuantumSimulation,SimulationWithSCCircuit,MakhlinReview}: (I) The
circuit can be manipulated by applying voltages, currents and/or magnetic
flux. Hence useful classical circuit techniques can be introduced in similar
ways. (II) Circuit manipulations can be implemented locally to a single
site/unit or globally to the whole system. (III) The circuit can be tailored
to certain characteristic frequency, interaction strength, etc., and the
circuit geometry can be fabricated in desired patterns. Furthermore, according
to recent reports the decoherence time of superconducting qubits based on
different superconducting circuits is approaching $0.1ms$
\cite{HPaikSchoelkopfPRL2011,Q factor 3D
Rigetti,ZKimWellstoodPalmerPRL2011,UQDP}. The Q factor of an on-chip
transmission line resonator \cite{Q factor 1D Martinis} can even go beyond
$10^{5}$. A 3D superconducting resonator \cite{Q factor 3D Schoelkopf, Q
factor 3D Rigetti} can have a quality factor up to $10^{9}$, which implies
that the life time of photons in superconducting resonators may approach
$10ms$. This is good enough to allow one to consider the photon number as a
conserved quantity in the circuit if compared to the operation frequency in
the circuit typically in the range of $100$MHz$-10$GHz
\cite{MakhlinReview,DevoretReview,DevoretReview2,ClarkeReview}.

Having those features of superconducting circuit in mind, we propose a scheme
to simulate the BHM with controllable inhomogeneous parameters. To demonstrate
some interesting features, we consider how the phase transition between the
delocalized SF and localized Mott insulator can be induced by manipulating the
parameters of one single site. In conventional setups, global parameters such
as the density or interaction drive the system across this transition, and
here we propose that in superconducting-circuit simulators, one may observe
this transition with a single-site manipulation. The details of our proposed
scheme are verified by the exact diagonalization method \cite{ED}, which already shows signatures of this transition in moderate-size
systems. Thus this proposed scheme should be feasible in experiments.

Here the simulator is based on an array of superconducting transmission line
resonators (TLRs). The goal is to simulate the BHM \cite{BHM}
\begin{equation}
H=-{\displaystyle\sum\limits_{i}}\mu_{i}n_{i}+{\displaystyle\sum\limits_{i}%
}\frac{U_{i}}{2}n_{i}(n_{i}-1)-{\displaystyle\sum\limits_{i}}t_{i}%
(b_{i}^{\dagger}b_{i+1}+b_{i}b_{i+1}^{\dagger}). \label{BoseHubbard}%
\end{equation}
Here $\mu_{i}$ is the on-site energy and it usually plays the role of the
chemical potential, $U_{i}$ is the on-site interaction, and $t_{i}$ is the
nearest-neighbor hopping coefficient. In cold atoms one can control the
filling and motion of a single atom \cite{GreinerNat}, but manipulations of
the energy and interaction on each site remain a challenge.

A superconducting TLR with a length in the range of centimeters can support a
microwave resonant frequency corresponding to the oscillations of the electric
potential and magnetic flux from the standing waves of the Cooper pair
density. Those microwaves are referred to as the photons in the TLR
\cite{CircuitQEDBlais}. The quantum electrodynamics (QED) framework can then
be applied to the TLR-qubit system to get the so-called circuit
QED\cite{CircuitQEDBlais}. A single site of the system is modeled by the
Jaynes-Cummings (JC) model \cite{JCM} while an array of circuit QED systems,
as schematically shown in Figure \ref{Circuit}, can be described by the
Jaynes-Cumming Hubbard model \cite{JCHM}%
\begin{align}
H  &  ={\displaystyle\sum\limits_{n}}[\hbar\omega_{n}^{c}a_{n}^{\dagger}%
a_{n}+\hbar\omega^{q}\sigma_{n}^{z}+g_{n}(a_{n}\sigma_{n}^{+}+a_{n}^{\dagger
}\sigma_{n}^{-})]\nonumber\\
&  +{\displaystyle\sum\limits_{n}}J_{n}(a_{n}^{\dagger}a_{n+1}+a_{n}%
a_{n+1}^{\dagger}),
\end{align}
where the parameters are $\omega_{n}^{c}$ as the cavity frequency, $\omega
^{q}$ as the qubit frequency, $g_{n}$ as the coupling strength between the
cavity and qubit, and $J_{n}$ as the hopping coefficient between cavities.

When the qubit is close to resonance with the cavity, they are co-excited and
the excitation on a single site has the form of a polariton. Simulating
polaritonic many body behavior has been studied recently based on various
physical systems \cite{cQED lattice,JCHM SC chain,BHM in Polaritons Chain}.
Here we consider a different regime in the parameter space to take advantage
of the tunability of superconducting quantum circuits. We focus on the
dispersive regime \cite{CQED BHM JC array}, where the excitation is limited in
the TLR while the qubit stays in its ground state. Hence the on-site
excitation becomes photonic. In this regime, a perturbation calculation shows
that the system can simulate the BHM. To make connections to experiments,
feasible controlling and probing methods of the quantum phase transition
between localized and delocalized states will be discussed. The exact
diagonalization (ED)\cite{ED} method is used to numerically demonstrate the details of 
the phase transition.

\section{Architecture of the simulator}

\label{secmodel} 
\begin{figure}[th]
\centering
\includegraphics[width=0.8\columnwidth]{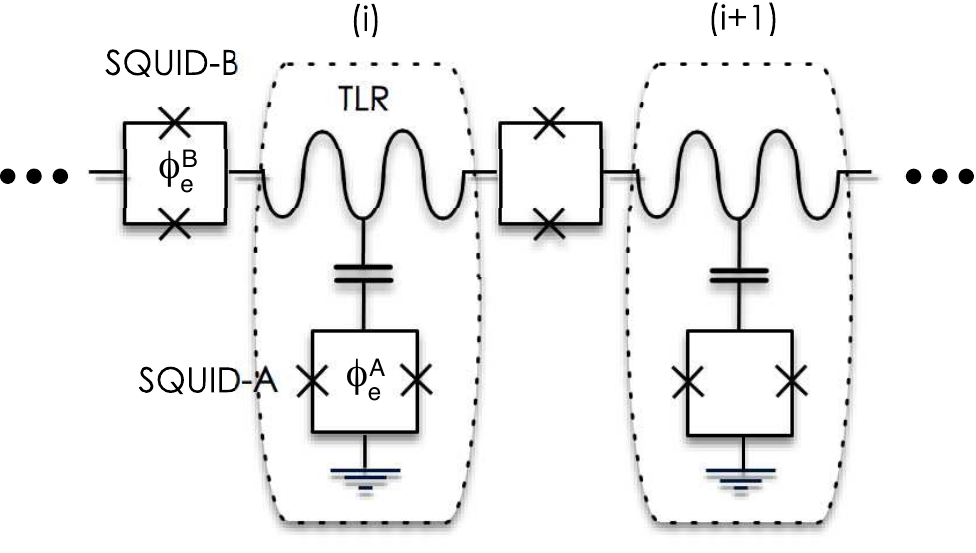}
\caption{Schematic plot of the 1D TLR array. SQUID-A as a tunable charge qubit
is capacitively coupled to the center of a TLR. Nearest neighbor sites are
connected by SQUID-B. The external magnetic flux $\phi_{e}^{A}$ and $\phi
_{e}^{B}$ through SQUID A and B can be used to tune their Josephson energies.
}%
\label{Circuit}%
\end{figure}

As illustrated in Figure \ref{Circuit}, the proposed
simulator is a one dimensional (1D) array of superconducting circuit elements.
One site is formed by a TLR capacitively coupled to a superconducting charge
qubit \cite{MakhlinReview,DevoretReview,DevoretReview2,ClarkeReview}, which is
labeled as SQUID-B, and the qubit energy is tunable. The TLRs on different
sites are connected via the SQUID-B, which leads to tunable couplings between
nearest neighbor sites. Here a derivation of how the Bose-Hubbard Hamiltonian
(\ref{BoseHubbard}) can be simulated by the superconducting circuit will be
presented. In order to simplify the derivation, we will use Hz$\times2\pi$ as
the unit of energy and set $\hbar\equiv1$. 

\subsection{TLR as a lattice element}

The qubit-TLR system is an analogue of an atom-cavity system. In quantum
optics the dynamics of the latter system can be modeled by the Janes-Cummings
Hamiltonian \cite{CircuitQEDBlais}. Our superconducting circuit Hamiltonian
can be derived following the work of circuit-QED in
Refs.~\cite{CircuitQEDBlais,CircuitQEDHouck,BishopSchoelkopfNPhys2009}. The
Hamiltonian of a single lattice site is
\begin{equation}
H^{site}=H^{TLR}+H^{qubit}.
\end{equation}
The TLR with length $D$ could be treated as a cavity with a single mode of the
first harmonic. Thus
\begin{align}
H^{TLR} &  =\frac{(2e)^{2}}{2C^{c}}N^{2}+\frac{1}{2L^{c}(2e)^{2}}(\phi
^{c})^{2}\nonumber\\
&  =\frac{1}{2}E_{c}^{c}N^{2}+\frac{1}{2}E_{L}^{c}(\phi^{c})^{2}\nonumber\\
&  =\omega^{c}a^{\dagger}a,\label{HTLR}%
\end{align}
where the cavity frequency $\omega^{c}=\frac{2\pi}{\sqrt{C^{c}L^{c}}}%
=2\pi\sqrt{E_{c}^{c}E_{L}^{c}}$, the net capacitance of the TLR is $C^{c}$,
the charge energy of the cavity $E_{c}^{c}=\frac{(2e)^{2}}{C^{c}}$, the net
inductance of the TLR is $L^{c}$ and after second quantization, the inductive
energy of the cavity is $E_{L}^{c}=\frac{1}{L^{c}(2e)^{2}}$. The node charge
number and node flux at the maximum points become
\begin{equation}
\left\{
\begin{array}
[c]{c}%
N=\sqrt{\omega^{c}/E_{c}^{c}}(a^{\dagger}+a)\\
\phi^{c}=-i\sqrt{\omega^{c}/E_{L}^{c}}(a^{\dagger}-a)
\end{array}
\right.  .
\end{equation}
For the first harmonic, the spatial distribution \cite{CircuitQEDBlais} of $N$
is $\cos(\frac{\pi}{D}x)$, $x\in\lbrack-\frac{D}{2},\frac{D}{2}]$, so the
maximum points are $x=-\frac{D}{2},0,\frac{D}{2}$ corresponding to the center
and two ends of the TLR. Since the qubit consists of two Josephson junctions
in a superconducting loop, the qubit Hamiltonian includes the capacitive
energy and inductive energy as
\begin{equation}
H^{qubit}=E_{c}^{A}(n-n_{g})^{2}+2E_{J}^{A}\cos(\frac{\phi_{e}^{A}}{2}%
)(1-\cos\phi).
\end{equation}
Here $n=C_{\Sigma}^{A}V_{J}/2e$ is the number of Cooper pairs on the island
and $n_{g}=C_{g}^{A}V_{g}/2e$ is the number of Cooper pairs on the gate, which
has a capacitance $C_{g}^{A}$ between the qubit and TLR. $E_{c}^{A}%
=\frac{(2e)^{2}}{2C_{\Sigma}^{A}}$ with $C_{\Sigma}^{A}$ being the total
effective capacitance in the qubit. The Josephson tunneling energy is
$E_{J}^{A}$ and the phase $\phi$ displaces the number of Cooper pairs. Casting
the Hamiltonian in Fock space and dropping the constant term $2E_{J}^{A}%
\cos(\frac{\phi_{e}^{A}}{2})$, one obtains
\begin{align}
H^{qubit} &  ={\displaystyle\sum\limits_{n}}[E_{c}^{A}(n-n_{g})^{2}\left\vert
n\right\rangle \left\langle n\right\vert \nonumber\\
&  +2E_{J}^{A}\cos(\frac{\phi_{e}^{A}}{2})(\left\vert n\right\rangle
\left\langle n+1\right\vert +\left\vert n+1\right\rangle \left\langle
n\right\vert )].
\end{align}
Because of the giant Kerr effect due to the Josephson junction, the energy
difference between the lowest two levels is separated from the other energies.
Therefore the SQUID-A can be considered as a superconducting qubit
\cite{MakhlinReview}, where the Pauli matrices are
\begin{align}
\widetilde{\sigma}^{x} &  =\left\vert 0\right\rangle \left\langle 1\right\vert
+\left\vert 1\right\rangle \left\langle 0\right\vert \\
\widetilde{\sigma}^{z} &  =-\left\vert 0\right\rangle \left\langle
0\right\vert +\left\vert 1\right\rangle \left\langle 1\right\vert .
\end{align}
Then we obtain
\begin{equation}
H^{qubit}=E_{c}^{A}\frac{1-2n_{g}}{2}\widetilde{\sigma}^{z}+2E_{J}^{A}%
\cos(\frac{\phi_{e}^{A}}{2})\widetilde{\sigma}^{x}.
\end{equation}
Here we have made use of
\begin{align}
{\displaystyle\sum\limits_{n}}(n-n_{g})^{2}\left\vert n\right\rangle
\left\langle n\right\vert  &  =n_{g}^{2}\left\vert 0\right\rangle \left\langle
0\right\vert +(1-2n_{g}+n_{g}^{2})\left\vert 1\right\rangle \left\langle
1\right\vert \nonumber\\
&  =\frac{1-2n_{g}}{2}\widetilde{\sigma}^{z}%
\end{align}
by dropping the constant term $(n_{g}^{2}+\frac{1-2n_{g}}{2})(\left\vert
0\right\rangle \left\langle 0\right\vert +\left\vert 1\right\rangle
\left\langle 1\right\vert )$. Hence the qubit Hamiltonian becomes a $2\times2$
matrix. The gate voltage $V_{g}$ is the electric potential at the point of the
TLR where the qubit couples to. This includes the DC gate voltage on the qubit
and a quantum mode of the TLR:
\begin{equation}
V_{g}=V^{dc}+\widehat{V^{ac}}.
\end{equation}
As Figure~\ref{Circuit} shows, the qubit is coupled to the center of the TLR
so the quantum mode of the voltage is
\begin{equation}
\widehat{V^{ac}}=\frac{2eN}{\sqrt{2}C^{c}}=V_{rms}(a^{\dagger}+a)
\end{equation}
for the fundamental mode, where $V_{rms}=\sqrt{{\omega^{c}/2C^{c}}}$ is the
root-mean-square value of the ground state voltage at the center of the TLR.
Hence
\begin{equation}
n_{g}=n^{dc}+C_{g}^{A}\sqrt{\omega^{c}/E_{c}^{c}}(a^{\dagger}+a).
\end{equation}
For the DC gate voltage bias at the degeneracy point $n^{dc}=\frac{1}{2}$,
\begin{equation}
H^{qubit}=E_{c}^{A}C_{g}^{A}\sqrt{\omega^{c}/E_{c}^{c}}(a^{\dagger
}+a)\widetilde{\sigma}^{z}+2E_{J}^{A}\cos(\frac{\phi_{e}^{A}}{2}%
)\widetilde{\sigma}^{x}.
\end{equation}
Using the qubit representation, we obtain
\begin{align}
\sigma^{x} &  =\left\vert \uparrow\right\rangle \left\langle \downarrow
\right\vert +\left\vert \downarrow\right\rangle \left\langle \uparrow
\right\vert \\
\sigma^{z} &  =-\left\vert \downarrow\right\rangle \left\langle \downarrow
\right\vert +\left\vert \uparrow\right\rangle \left\langle \uparrow
\right\vert
\end{align}
While biased at the degeneracy point, the eigenbasis of the qubit Hamiltonian
is given by $\left\vert \uparrow\right\rangle =(\left\vert 0\right\rangle
+\left\vert 1\right\rangle )/\sqrt{2}$ and $\left\vert \downarrow\right\rangle
=(\left\vert 0\right\rangle -\left\vert 1\right\rangle )/\sqrt{2}$. The
combined system for one site now has the following form
\begin{align}
H^{site} &  =2e\frac{C_{g}^{A}}{C_{\Sigma}^{A}}\sqrt{\omega^{c}C^{c}%
}(a^{\dagger}+a)\sigma^{x}+\frac{\omega^{q}}{2}\sigma^{z}+\omega^{c}%
a^{\dagger}a\nonumber\\
&  =\omega^{c}a^{\dagger}a+\frac{\omega^{q}}{2}\sigma^{z}+g^{q}\sigma
^{x}(a^{\dagger}+a),
\end{align}
where $\omega^{q}=4E_{J}^{A}\cos(\frac{\phi_{e}^{A}}{2}).$

The magnitudes of the qubit frequency and cavity frequency are in the same
range of about $10$GHz, so it is natural to apply the rotating wave
approximation (RWA). Let $\Delta=\omega^{c}-\omega^{q}$ denote the detuning
between the cavity and qubit frequencies. Then $\Delta=\omega^{c}-\omega
^{q}\ll\omega^{c}+\omega^{q}$. Moving into the interaction picture and
rotating frame one gets the Jaynes-Cummings interaction%
\begin{align}
H_{int}^{rot}  &  =g^{q}(\sigma_{+}e^{i\omega^{q}t}+\sigma_{-}e^{-i\omega
^{q}t})(a^{\dagger}e^{i\omega^{c}t}+ae^{-i\omega^{c}t})\nonumber\\
&  \overset{RWA}{\approx}g^{q}(\sigma_{-}a^{\dagger}e^{i\Delta t}+\sigma
_{+}ae^{-i\Delta t}),
\end{align}
where the fast oscillation terms with the phase $e^{i(\omega^{c}+\omega^{q}%
)t}$ and $e^{-i(\omega^{c}+\omega^{q})t}$ are neglected in the RWA.  Moving
back to the non-rotating frame we get the JC Hamiltonian%
\begin{align}
H^{site}  &  =\omega^{c}a^{\dagger}a+\frac{\omega^{q}}{2}\sigma_{z}%
+g^{q}(\sigma_{-}a^{\dagger}+\sigma_{+}a)\\
&  =H_{0}+V,\nonumber
\end{align}
where the diagonal term is $H_{0}=\omega^{c}(a^{\dagger}a+\sigma^{z})$ and the
interaction term is $V=\Delta\sigma^{z}/2+g^{q}(\sigma_{-}a^{\dagger}%
+\sigma_{+}a)$. Here we consider the dispersive regime \cite{CQED BHM JC
array,FWMtoolbox} so $\Delta\gg g^{q}$ and there is no excitation from
$\left\vert g\right\rangle $ to $\left\vert e\right\rangle $. Moreover,
$g^{q}(\sigma^{-}a^{\dagger}+\sigma^{+}a)$ becomes a perturbation term. In
order to get higher-order effective interactions we apply the standard
perturbation theory to the fourth order and obtain
\[
E_{g}^{(0)}=0,E_{e}^{(0)}=\Delta, V_{gg}=V_{ee}=0,V_{ge}=g^{q}a^{\dagger
}=V_{ge}^{\dagger}.
\]
Hence we only consider the correction terms for $E_{g}^{(0)}.$%
\begin{align}
E_{g}^{(1)}  &  =V_{gg}=0,\\
E_{g}^{(2)}  &  =-\frac{g^{q2}}{\Delta}aa^{\dagger},\\
E_{g}^{(3)}  &  =\frac{V_{ge}V_{ee}V_{eg}}{\Delta^{2}}=0,\\
E_{g}^{(4)}  &  =(\frac{g^{q}}{\Delta})^{3}g^{q}a^{\dagger}aa^{\dagger}a.
\end{align}
Then%
\begin{equation}
V=-\frac{g^{q2}}{\Delta}aa^{\dagger}+(\frac{g^{q}}{\Delta})^{3}g^{q}%
a^{\dagger}aa^{\dagger}a.
\end{equation}
Here the Kerr term $(\frac{g^{q}}{\Delta})^{3}g^{q}a^{\dagger}aa^{\dagger}a$
gives rise to an effective on-site interaction. Going back to the Schrodinger
picture, the single-site Hamiltonian becomes%
\begin{align}
H^{site}  &  =(\omega^{c}-\frac{g^{q2}}{\Delta}+(\frac{g^{q}}{\Delta}%
)^{3}g^{q})a^{\dagger}a\label{Hsite}\\
&  +\frac{\omega^{q}}{2}\sigma^{z}+(\frac{g^{q}}{\Delta})^{3}g^{q}a^{\dagger
}a(a^{\dagger}a-1).\nonumber
\end{align}
The charge qubit could be either a single Cooper-pair transistor (SCT) or a
transmon
\cite{MakhlinReview,ClarkeReview,DevoretReview,KochSchoelkopfPRA2007TransmonQubit}
whose qubit frequency can be tuned by changing the magnetic flux bias through
a SQUID loop in the qubit circuit. The detuning $\Delta$ is a
controllable parameter. $\omega^{c,eff}=\omega^{c}-\frac{g^{q2}}{\Delta
}+(\frac{g^{q}}{\Delta})^{3}g^{q}$ is the effective cavity frequency and
$U=2(\frac{g^{q}}{\Delta})^{3}g^{q}$ becomes the effective on-site interaction
energy of the photons. Both of them are functions of $\Delta$. Assuming
$g^{q}=120$MHz$\times2\pi$\cite{DevoretReview2,ClarkeReview}, $\Delta\geq
0.9$GHz$\times2\pi$ so $\left(  \omega^{c}-\omega^{c,eff}\right)  \in\left[
-0.1,0.1\right]  $GHz$\times2\pi$. We remark that the case $\Delta\sim g^{q}$,
where the excitations are polaritons rather than photons, has been discussed
in the literature \cite{BHM in Polaritons Chain}.

\subsection{Tunable TLR array}

Tunable couplings between different sites are necessary in simulating the BHM.
Different architectures for implementing a tunable coupler between two
superconducting TLRs have been realized and discussed in
Refs.~\onlinecite{CPSun
Tunable Coupler, BorjaTunable, FastTunableCoupler,
YuChenTunableCoupler,JQLiaoPRA2009,PintoPRB2010,SrinivasanHouckPRL2011TunableCupling}.
Here we present a basic design as shown in Figure \ref{Circuit} to demonstrate
our quantum simulator. SQUID B with different size and energy from those of
SQUID A is used to couple the TLRs. The coupling term is from SQUID B and
\begin{equation}
H^{B}={\displaystyle\sum\limits_{i=upp,low}}[\frac{C_{J}^{B}}{2}\left(
\dot{\phi}_{i}^{jj}\right)  ^{2}+E_{J}^{B}(1-\cos\phi_{i}^{jj})],
\end{equation}
where $\phi_{i=upp,low}^{jj}$ are the phase differences across the upper and
lower Josephson junctions of SQUID B (see Fig.~\ref{Circuit}). By changing of variables $\phi_{e}%
^{B}=\phi_{upp}^{jj}+\phi_{low}^{jj},$ where $\phi_{e}^{B}$ is the external
magnetic flux bias through SQUID B, $\dot{\phi}_{upp}^{jj}+\dot{\phi}%
_{low}^{jj}=\dot{\phi}_{e}^{B}=0$. Let the node phases on the two ends that
connect to TLR 1 and 2 be $\phi_{1}^{c}$ and $\phi_{2}^{c}$. According to the
geometry of the SQUIDs, $\phi_{1}^{c}-\phi_{2}^{c}=\frac{1}{2}($ $\phi
_{upp}^{jj}-\phi_{low}^{jj})$ so $\dot{\phi}_{upp}^{jj}-\dot{\phi}_{low}%
^{jj}=2(\dot{\phi}_{1}^{c}-\dot{\phi}_{2}^{c})$. After some algebra, one gets
$\left(  \dot{\phi}_{upp}^{jj}\right)  ^{2}+\left(  \dot{\phi}_{low}%
^{jj}\right)  ^{2}=2(\dot{\phi}_{1}^{c})^{2}-4\dot{\phi}_{1}^{c}\dot{\phi}%
_{2}^{c}+2(\dot{\phi}_{2}^{c})^{2}$. Here we define $N_{1,2}$ as the number of
Cooper pairs on the node connected to TLR 1 or 2, so $\frac{C_{J}^{B}}%
{2}\left(  \dot{\phi}_{1,2}^{c}\right)  ^{2}=\frac{1}{2}\frac{(2e)^{2}}%
{C_{J}^{B}}N_{1,2}^{2}=E_{c}^{B}N_{1,2}^{2}$. Therefore the charge energy
becomes $(i=upp,low)$%
\begin{align}
{\sum\limits_{i}}\frac{C_{J}^{B}}{2}\left(  \dot{\phi}_{i}^{jj}\right)  ^{2}
&  =C_{J}^{B}(\overset{\cdot}{\phi}_{1}^{c})^{2}-2C_{J}^{B}\overset{\cdot
}{\phi}_{1}^{c}\overset{\cdot}{\phi}_{2}^{c}+C_{J}^{B}(\overset{\cdot}{\phi
}_{2}^{c})^{2}\nonumber\\
&  =2E_{c}^{B}N_{1}^{2}-4E_{J}^{B}N_{1}N_{2}+2E_{c}^{B}N_{2}^{2}%
.\label{chargeH}%
\end{align}
We also assume that the two Josephson junctions in SQUID B are uniform. By
neglecting some constant terms, the Josephson energy becomes ($i=upp,low$)
\begin{align}
{\sum\limits_{i}}E_{J}^{B}(1-\cos\phi_{i}^{jj}) &  =-2E_{J}^{B}\cos(\frac
{\phi_{e}^{B}}{2})\cos(\frac{\phi_{upp}^{jj}-\phi_{low}^{jj}}{2})\nonumber\\
&  =-2E_{J}^{B}\cos(\frac{\phi_{e}^{B}}{2})\cos(\phi_{1}^{c}-\phi_{2}^{c}).
\end{align}
It will be shown that $2E_{J}^{B}\cos(\frac{\phi_{e}^{B}}{2})$ can be tuned to the
same order of magnitude as the on-site interaction term $(\frac{g^{q}}{\Delta
})^{3}g^{q}$ in Eq.~(\ref{Hsite}), which is needed to place the system near the
MI-SF phase transition. Moreover, the phase difference $\left(  \phi
_{upp}^{jj}-\phi_{low}^{jj}\right)  $ can initially be set to zero by shorting
both sides. Expanding $\cos(\frac{\phi_{upp}^{jj}-\phi_{low}^{jj}}{2})$ to the
second order, one obtains ($i=upp,low$)
\begin{equation}
{\sum\limits_{i}}E_{J}^{B}(1-\cos\phi_{i}^{jj})\simeq E_{J}^{B}\cos(\frac
{\phi_{e}^{B}}{2})[(\phi_{1}^{c})^{2}-2\phi_{1}^{c}\phi_{2}^{c}+(\phi_{2}%
^{c})^{2}].\label{JosephsonH}%
\end{equation}

Combining Eqs. (\ref{chargeH}) and (\ref{JosephsonH}), one gets the
Hamiltonian for SQUID B
\begin{align}
H^{B} &  ={\sum\limits_{i=1,2}[2E_{c}^{B}N_{i}^{2}+E_{J}^{B}\cos(\frac
{\phi_{e}^{B}}{2})(\phi_{i}^{c})^{2}]}\nonumber\\
&  -[4E_{J}^{B}N_{1}N_{2}+2E_{J}^{B}\cos(\frac{\phi_{e}^{B}}{2})\phi_{1}%
^{c}\phi_{2}^{c}]\\
&  =H_{1,2}^{TLR^{\prime}}+H^{coup}.
\end{align}
Here the simple harmonic terms $H_{1,2}^{TLR^{\prime}}$ give additional
frequency shift to the TLR Hamiltonian in Eq. (\ref{HTLR}). Since the net TLR
Hamiltonian is
\begin{align}
{H}_{net,i}^{TLR} &  =\frac{1}{2}(E_{c}^{c}+{4E_{c}^{B})N}_{i}^{2}+\frac{1}%
{2}[E_{L}^{c}+2{E_{J}^{B}\cos(\frac{\phi_{e}^{B}}{2})](\phi_{i}^{c})^{2}%
}\nonumber\\
&  =\frac{1}{2}E_{c}^{c\ast}{N}_{i}^{2}+\frac{1}{2}E_{L}^{c\ast}{(\phi_{i}%
^{c})^{2},}%
\end{align}
the dressed cavity frequency becomes%
\begin{equation}
{\omega}^{c\ast}=2\pi\sqrt{E_{c}^{c\ast}E_{L}^{c\ast}}.
\end{equation}
Once the TLRs are connected into an array with those SQUID Bs, the fundamental
cavity frequency changes from $\omega^{c}$ to $\omega^{c\ast}$. Moreover, TLR
1 and 2 are coupled by
\begin{align}
H^{coup} &  =-[4E_{J}^{B}N_{1}N_{2}+2E_{J}^{B}\cos(\frac{\phi_{e}^{B}}{2}%
)\phi_{1}^{c}\phi_{2}^{c}]\nonumber\\
&  =-g^{cap}(a_{1}^{\dagger}+a_{1})(a_{2}^{\dagger}+a_{2})\\
&  +g^{ind}(a_{1}^{\dagger}-a_{1})(a_{2}^{\dagger}-a_{2}).
\end{align}
Here the coupling constants are%
\begin{equation}
\left\{
\begin{array}
[c]{c}%
g^{cap}=\frac{\omega^{c}E_{c}^{B}}{E_{c}^{c\ast}}\\
g^{ind}=\frac{\omega^{c}4E_{J}^{B}\cos(\frac{\phi_{e}^{B}}{2})}{E_{L}^{c\ast}}%
\end{array}
\right.  .
\end{equation}
A similar coupling Hamiltonian can be found in Ref.~\onlinecite{BorjaTunable},
which is supported by experiments \cite{GrossTunableExper}. By considering two
identical resonators $\omega_{1}^{c\ast}=\omega_{2}^{c\ast}$ and applying the
RWA and conservation of the photon number, one obtains
\[
H^{coup}\simeq-(g^{cap}+g^{ind})(a_{1}^{\dagger}a_{2}+a_{1}a_{2}^{\dagger
})=g(a_{1}^{\dagger}a_{2}+a_{1}a_{2}^{\dagger}).
\]
The TLR-SQUID-TLR (TST) system has the Hamiltonian%
\begin{equation}
H^{TST}={\displaystyle\sum\limits_{i=1,2}}\hbar\omega_{i}^{c\ast}%
a_{i}^{\dagger}a_{i}-g(a_{1}^{\dagger}a_{2}+a_{1}a_{2}^{\dagger}).
\end{equation}

We consider typical values \cite{DevoretReview2,ClarkeReview} of $E_{c}%
^{B}=300$MHz$\times2\pi$, $E_{J}^{B}=500$MHz$\times2\pi$, $E_{c}^{c\ast}%
=10$GHz$\times2\pi$, $E_{L}^{c\ast}=10$GHz$\times2\pi$. Note that $\phi
_{e}^{B}$ can be tuned within $[0,2\pi]$, so $g^{ind}\in\lbrack2,-2]$%
GHz$\times2\pi$. The net coupling strength is $g=-(g^{cap}+g^{ind})\in
\lbrack-2.3,1.7]$GHz$\times2\pi$. Since the perturbation approach is applied
to the on-site Hamiltonian, in order to keep $H^{coup}$ with the same order of
magnitude as the highest order term in Eq.~(\ref{Hsite}), the coupling
strength $g$ has to fulfill the condition $g<g^{q}$. By biasing the system in
the range $\phi_{e}^{B}$ around $\pi$, one should be able to get a smaller
range of $g\in\lbrack-30,30]$MHz$\times2\pi$.

\subsection{Superconducting-circuit simulator of the BHM}
Combining the on-site Hamiltonian and couplings between nearest neighbor
sites, we obtain a many-body Jaynes-Cumming Hubbard Hamiltonian:%
\begin{align}
H^{JCHM}  &  ={\displaystyle\sum\limits_{i=1}^{N}}[\hbar\omega_{i}^{c\ast
}-\frac{g_{i}^{q2}}{\Delta}+(\frac{g_{i}^{q}}{\Delta})^{3}g_{i}^{q}%
]a_{i}^{\dagger}a_{i}+{\displaystyle\sum\limits_{i=1}^{N}}\hbar\omega_{i}%
^{q}\sigma_{i}^{z}\nonumber\\
&  +{\displaystyle\sum\limits_{i=1}^{N}}(\frac{g_{i}^{q}}{\Delta})^{3}%
g_{i}^{q}a_{i}^{\dagger}a_{i}(a_{i}^{\dagger}a_{i}-1)\nonumber\\
&  -{\displaystyle\sum\limits_{i=1}^{N-1}}(g_{i}^{cap}+g_{i}^{ind}%
)(a_{i}^{\dagger}a_{i+1}+a_{i}a_{i+1}^{\dagger}) \label{JCHHamiltonian}%
\end{align}
In the dispersive regime, where our perturbation approach is applicable, the
qubit does not get excitations and stays in its ground state. Therefore the
qubit term ${\displaystyle\sum\limits_{i=1}^{N}}\omega_{i}^{q}\sigma_{i}^{z}$
does not contribute to the many-body energy. In this case, the Jaynes-Cummings
lattice model can be mapped to the Bose Hubbard model\cite{CQED BHM JC array}
by neglecting the qubit term from Eq.~(\ref{JCHHamiltonian}) and treating the
photons in the TLR as interacting bosons. 

When compared to
Eq.~(\ref{BoseHubbard}), the on-site energy, on-site interaction, and hopping
terms are
\begin{align}
\mu_{i}  &  =-[\omega_{i}^{c\ast}-(\frac{g_{i}^{q}}{\Delta_{i}})g_{i}%
^{q}+(\frac{g_{i}^{q}}{\Delta_{i}})^{3}g_{i}^{q}]\\
\frac{U_{i}}{2}  &  =(\frac{g_{i}^{q}}{\Delta_{i}})^{3}g_{i}^{q}%
\label{Onsite Interaction}\\
t_{i}  &  =(g_{i}^{cap}+g_{i}^{ind})=g_{i}.
\end{align}
As discussed previously, $\Delta_{i}$ and $g_{i}$ can be tuned by a magnetic
flux bias, so they are the independent variables in this model. One may recall
that $\left\vert t\right\vert =\left\vert g\right\vert \in\lbrack
0,30]$MHz$\times2\pi$ from previous discussions. In the dispersive regime
$\left\vert \Delta\right\vert \in\lbrack0.9,1.2]$GHz$\times2\pi$ should give
reasonable values \cite{DevoretReview2,ClarkeReview} of $g^{q}=120$%
MHz$\times2\pi$. Thus $g^{q}/t\in\lbrack4,+\infty)$, $\left\vert
\Delta/t\right\vert \in\lbrack30,+\infty)$. In terms of the BHM,
$U/t\in(0,+\infty)$, which implies that the range of $U/t$ that can be
simulated by this simulator should cover the MI-SF transition. To avoid going
beyond the valid range of our approximation, the parameters are chosen in the
range $\left\vert \Delta/t\right\vert \in\lbrack30,10^{3}]$.

In this simulation scheme one may notice that the on-site energy $\mu_{i}$,
interaction strength $U_{i}$, and hopping coefficient $t_{i\text{ }}$ can be
explicitly made site-dependent. Therefore, this superconducting TLR array can
be a versatile simulator of the BHM, especially if phenomena due to spatial
inhomogeneity are of interest. Furthermore, compared to ultracold atoms in
optical lattices, this superconducting circuit simulator has some additional
features. As we already emphasized, all parameters can be tuned individually
and this makes it possible to study problems in various geometries. Moreover,
the interacting bosons in the simulator is confined inside the TLRs so there
is no need for background trapping potentials, which is common in cold-atom
systems. Moreover, open boundary conditions (OBCs) with hard walls can be
introduced by terminating the coupling SQUID at the ends of the
superconducting TLR array. Even though weak capacitive couplings to the leads
at the two ends of the array may be present, a high Q factor can still be
maintained \cite{Q factor 1D Martinis}. On the other hand, periodic boundary
conditions (PBCs) can be realized by fabricating a 1D array into a loop
structure. Hence bulk properties can be studied with a small number of sites
with minimal boundary effects. The examples given in the following sections
illustrates those features of the superconducting circuit simulator.

\section{Single-site manipulations of the MI-SF transition}

\label{sec_numerical} 
\begin{figure}[th]
\centering
\includegraphics[width=0.9\columnwidth]{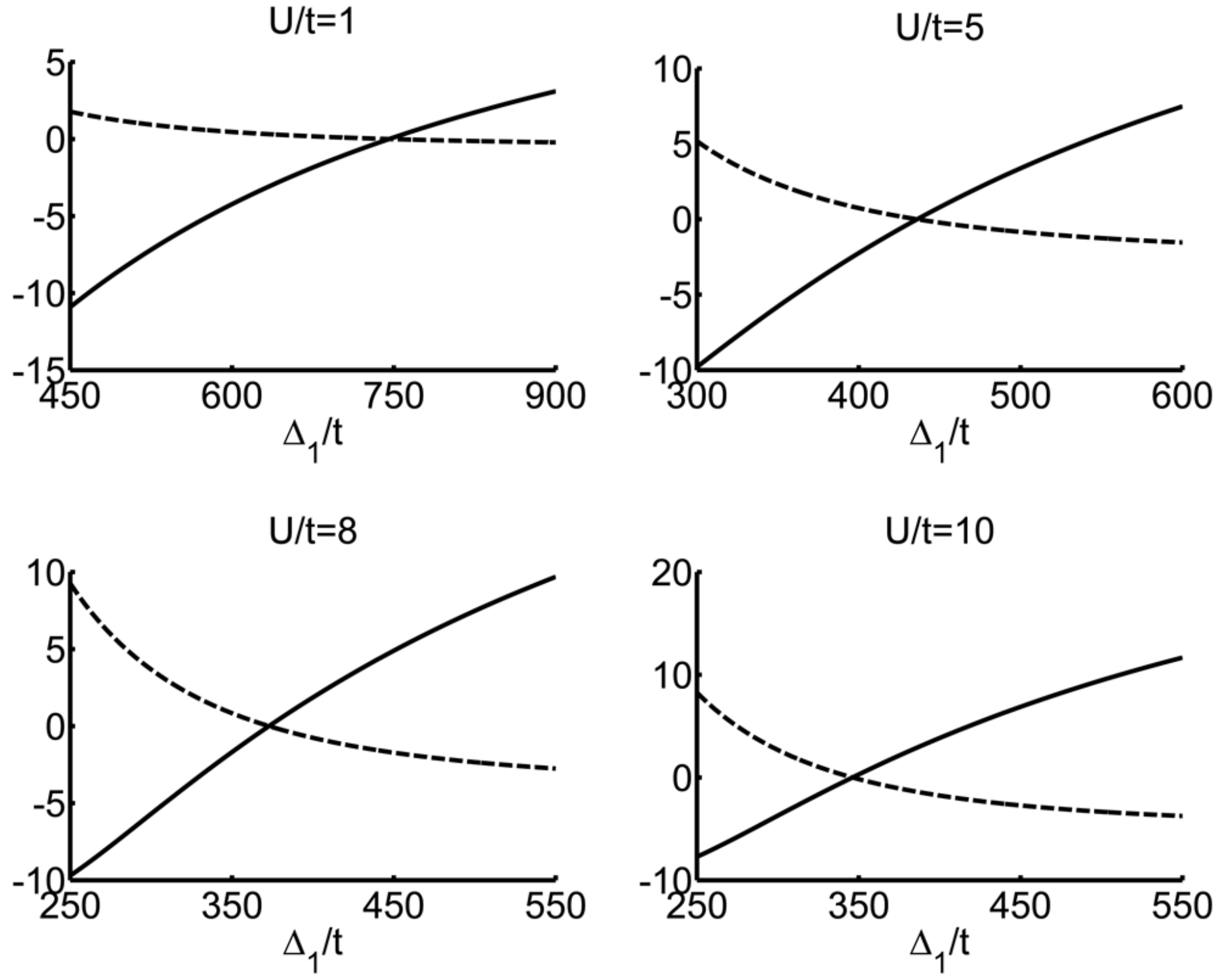}\caption{$\delta$
(solid lines) and $\eta$ (dashed lines) as functions of $\Delta_{1}$ for
$U/t=1,5,8,10$ and $g^{q}=120$MHz$\times2\pi$. As Eq.~(\ref{Site wise}) shows,
$\delta$ and $\eta$ are the displacements of the on-site energy and on-site
interaction of the first site.}%
\label{Energy}%
\end{figure}

Here we present one interesting application of this
superconducting circuit simulator, where the MI-SF transition of the BHM can
be induced by single-site manipulations. Other possible applications will be
discussed later. To concentrate on the underlying physics, we consider a 1D
array of $N$ sites. For simplicity, the parameters of a selected site (called
site 1) is tuned by external magnetic flux through the charge qubit coupled to
the TLR of this site. One may consider, for site 1, a shift of the onsite
energy by $\delta$ and a shift of the onsite coupling constant by $\eta$. The
choice of which site should be manipulated is not important since the
conclusions remain the same for the case with PBC. From the BHM
(\ref{BoseHubbard}), the Hamiltonian of this 1D array is rewritten in the
form
\begin{align}
H  &  =\delta n_{1}+\eta n_{1}(n_{1}-1)-\mu{\displaystyle\sum\limits_{i=1}%
^{N}}n_{i}\nonumber\\
&  +\frac{U}{2}\sum_{i=1}^{N}n_{i}(n_{i}-1)-t\sum_{i}^{N^{\prime}}%
(b_{i}^{\dagger}b_{i+1}+b_{i+1}^{\dagger}b_{i}), \label{Site wise}%
\end{align}
where
\begin{equation}
\left\{
\begin{array}
[c]{c}%
\delta=-g_{q}^{2}(\frac{1}{\Delta_{1}}-\frac{1}{\Delta_{i}})+g_{q}^{4}\left[
(\frac{1}{\Delta_{1}})^{3}-(\frac{1}{\Delta_{i}})^{3}\right] \\
\eta=g_{q}^{4}\left[  (\frac{1}{\Delta_{1}})^{3}-(\frac{1}{\Delta_{i}}%
)^{3}\right]
\end{array}
\right.  \text{ .} \label{Varables}%
\end{equation}
Here $\Delta_{1}$ is the detuning energy between the qubit and TLR on the site
1 while $\Delta_{i}$ is the detuning of the other sites. A diagram of $\delta$
and $\eta$ as a function of $\Delta_{1}$ is shown in Figure~\ref{Energy},
which gives an estimation of the BHM parameters in the presence of a
single-site manipulation. 

\begin{figure}[ptb]
\centering
\includegraphics[width=0.7\columnwidth]{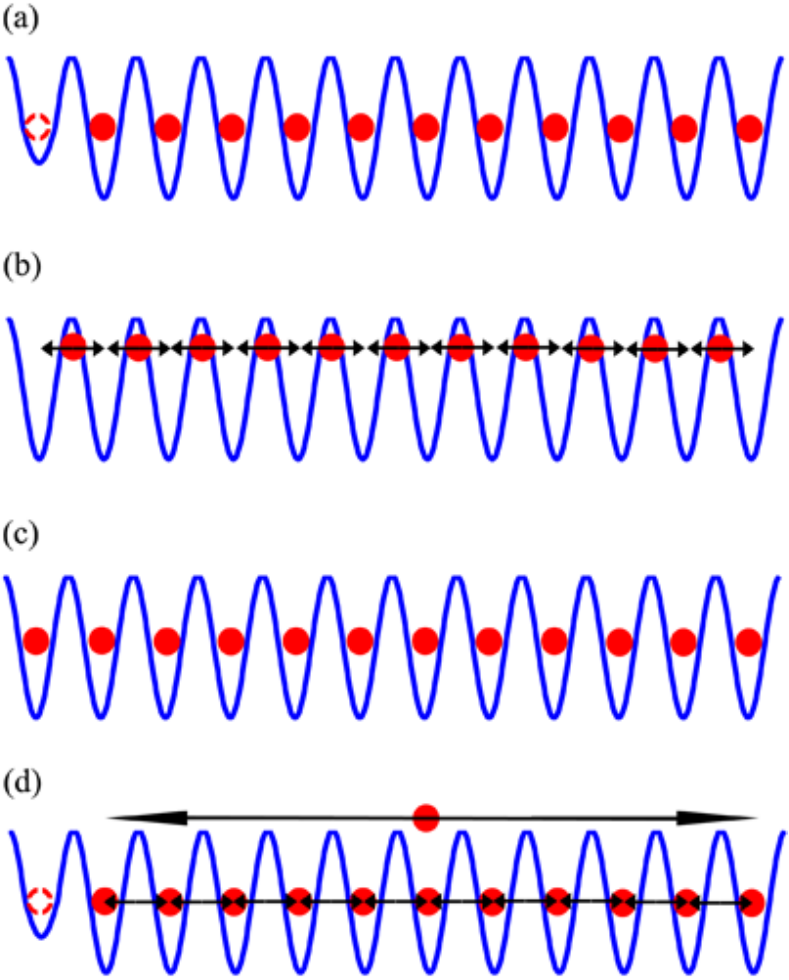}\caption{ (a) and
(b) illustrate the Mott insulator to superfluid transition for $N-1$ bosons
with strong repulsion in $N$ sites. (a) The on-site energy of site-1 is
increased and this pushes the system into a localized Mott insulator phase.
The dashed circle means the first site is virtually empty due to its large
on-site energy. (b) The system becomes a delocalized superfluid state when the
on-site energy is about the same as that of the other sites. (c) and (d)
illustrate the transition for $N$ bosons with strong repulsion in $N$ sites.
(c) When the array is uniform, the bosons are in a localized Mott insulator
phase. (d) By increasing the on-site energy of site 1, photons are pushed into
the bulk and form a delocalized superfluid.}%
\label{Cases}%
\end{figure}

In the upper limit of the summation, $N^{\prime}=N-1$ is for the OBC while
$N^{\prime}=N$ is for the PBC. We keep $t_{i}=t$ the same in the whole lattice
because it does not depend on $\Delta_{1}$. We vary $\Delta_{1}/t$ as an
independent variable. The unit of energy will be $t$. The advantages of this
protocol are: (1) The qubit energy is intact away from the manipulated site.
(2) Particles are conserved in the whole system. We define the particle
density $\rho$ as the ratio between the photon number and site number. In the
following we consider the phase transition due to this single-site
manipulation when $\rho<1$ and $\rho=1$. For $\rho=(N-1)/N$ the system is a
delocalized SF state in the absence of manipulations and a single-site push
leads to a localized MI state, which is shown schematically in
Fig.~\ref{Cases}(a)(b). The second case with $\rho=1$ is illustrated by
Fig.~\ref{Cases}(c)(d), where the system is in an MI state without
manipulations and becomes an SF after a single-site push.

To characterize those single-site manipulated transitions and to identify
where the transitions take place, we analyze a useful quantity called the
fidelity metric, which has been shown to capture quantum phase transitions or
sharp quantum crossovers in fermion Hubbard model \cite{JiaFidelity,
Fidelity_Hubbard_analytical} and other model Hamiltonians \cite{Fidelity
Metric, FidelityScaling}. Given a Hamiltonian of the form $H(\lambda
)=H_{0}+\lambda H_{1}$, the fidelity is defined as the overlap between two
(renormalized) ground states obtained with a small change $\delta\lambda$ in
the parameter $\lambda$:
\begin{equation}
F(\lambda,\delta\lambda)=\langle\Phi_{0}(\lambda)|\Phi_{0}(\lambda
+\delta\lambda)\rangle.
\end{equation}
However, the fidelity has been shown to be an extensive quantity that scales
with the system size \cite{Fidelity_on_Hubbard_Model,FidelityScaling}.
Therefore, the fidelity metric is induced as
\cite{fidelityorigin2,Fidelity_Hubbard_analytical,FidelityScaling}
\begin{equation}
g(\lambda,\delta\lambda)=(2/N)(1-F(\lambda,\delta\lambda))/\delta\lambda^{2},
\end{equation}
whose limit as $\delta\lambda\rightarrow0$ is well defined away from the
critical points and standard perturbation theories apply. More precisely,
\begin{equation}
\lim_{\delta\lambda\rightarrow0}g(\lambda,\delta\lambda)=\frac{1}{N}%
\sum_{\alpha\neq0}\frac{|\langle\Phi_{\alpha}(\lambda)|H_{1}|\Phi_{0}%
(\lambda)\rangle|^{2}}{[E_{0}(\lambda)-E_{\alpha}(\lambda)]^{2}}.
\end{equation}
The fidelity metric measures how significantly the ground-state wave function
changes as the parameter $\lambda$ changes. A dramatic increase of the
fidelity metric as a function of the varying parameter indicates a quantum
phase transition or sharp quantum crossover \cite{Fidelity Metric}.

\subsection{Case 1: $\rho<1$}

\begin{figure}[th]
\centering
\includegraphics[width=1.\columnwidth]{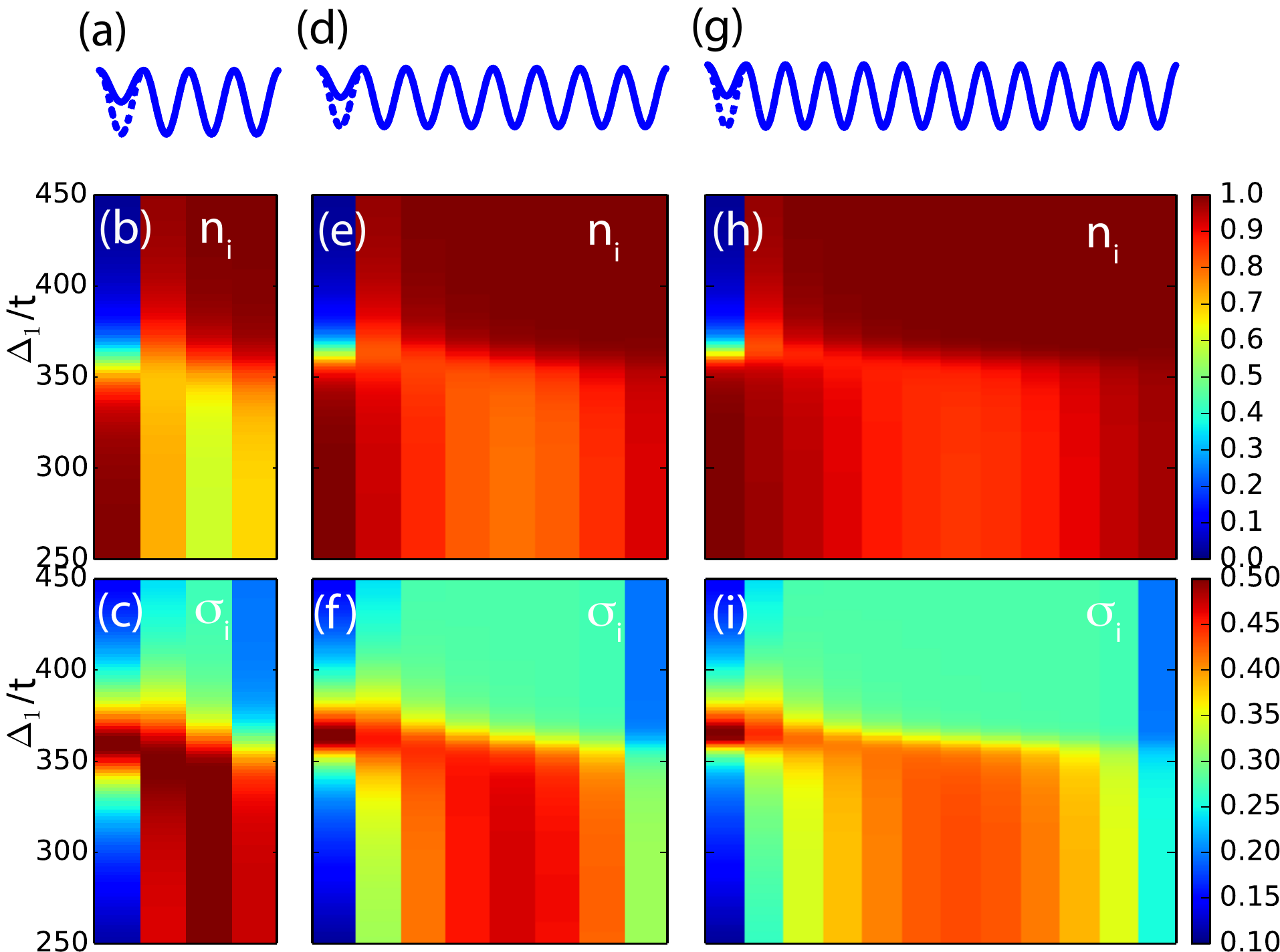}\caption{Exact
diagonalization results of the density $n_{i}$ and its variance $\sigma_{i}$
as a functions of $\Delta_{1}$ for Case-1 with OBC. Site $2$ to $N$ are
uniform and $U=10t$. (a)-(c) show the results for a $4$-site array with $3$
photons. In (a) the dashed line and solid line on the first site correspond to
the two schemes shown in Fig. \ref{Cases}. (d)-(f) correspond to the case of
$8$ sites with $7$ photons. (g)-(i) correspond to $12$ sites with $11$
photons. }%
\label{Ground state}%
\end{figure}

\begin{figure*}[th]
\centering
\includegraphics[width=2\columnwidth]{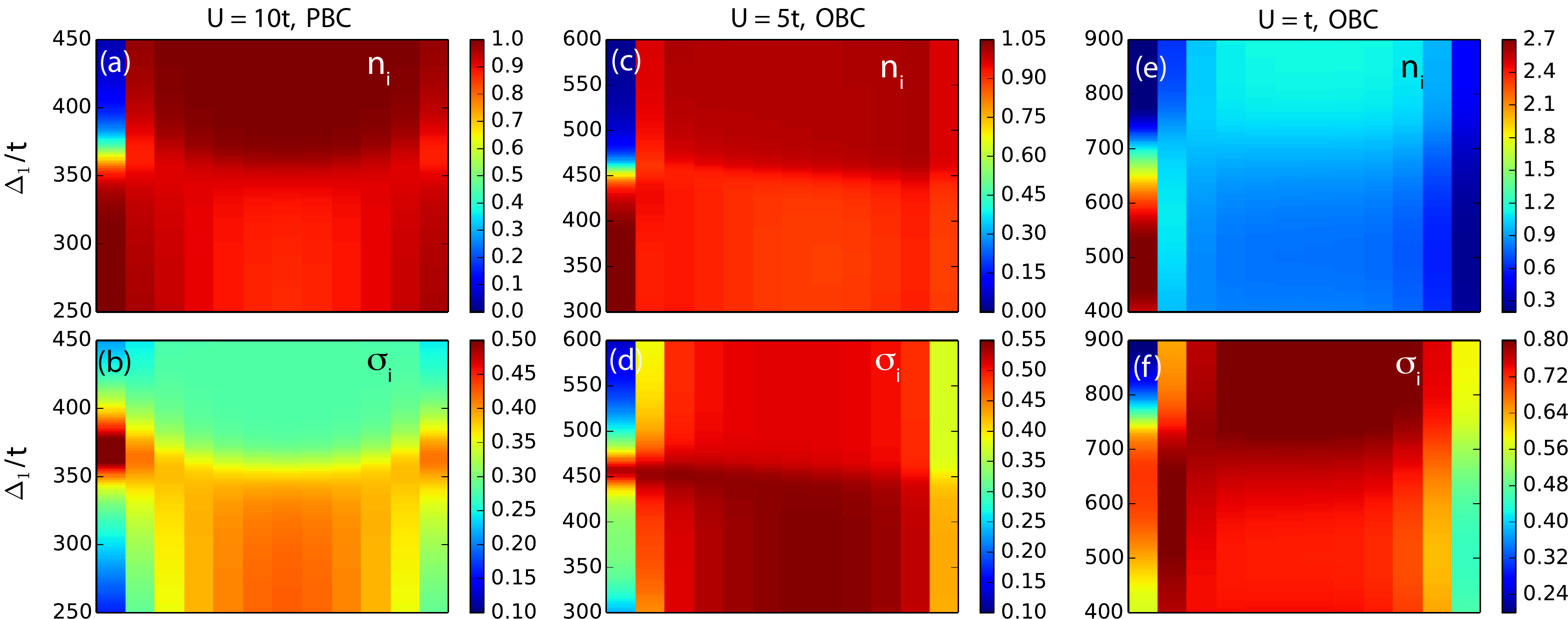}\caption{Photon
density profiles and its variance for selected values of $U$ and boundary
conditions. (a) and (b): $U/t=10$ and PBC. In this case, the photons in site
$2$ and $N$ can both tunnel to site $1$. Hence the photon density on site $2$
and $N$ are different from the bulk value due to boundary effects. (c) and
(d): $U/t=5$ and OBC. (e) and (f): $U/t=1$ and OBC. The non-uniform density
and its significant variance of the last case indicate that there is no Mott
insulator in this setting. Here $N=12$ with $11$ photons. }%
\label{GroundstateCaseI}%
\end{figure*}

When there are $(N-1)$ photons in an array of $N$ sites, the ground state
should be delocalized due to the incommensurate filling if all the sites have
the same on-site energy and interaction energy. As will be shown in Figure
\ref{Ground state} and Figure~\ref{GroundstateCaseI}, non-uniform
distributions of $n_{i}$ and stronger fluctuations of the on-site photon
density, quantified by the variance $\sigma_{i}=\left\langle \left\langle
n_{i}^{2}\right\rangle -\left\langle n_{i}\right\rangle ^{2}\right\rangle $,
in the small $\Delta_{1}$ regime indicates delocalization of the photons with
interactions up to $U=10t$. By increasing the on-site energy of site 1, which
can be performed by increasing $\Delta_{1}$, a transition to a localized MI
state of the remaining $N-1$ sites occurs. The setup is summarized in Figure
\ref{Cases}(a)(b). Based on current experimental technology \cite{cQEDarray
Expr, SimulationWithSCCircuit,MooijSCNetwork, FazioQuanPhasTSCCircuit}, the
size of the lattice in our exact diagonalization are chosen as $N=4,8,12$. An
estimation of the phase transition point can be obtained from a mean-field
approximation. 

\begin{figure}[th]
\centering
\includegraphics[width=\columnwidth]{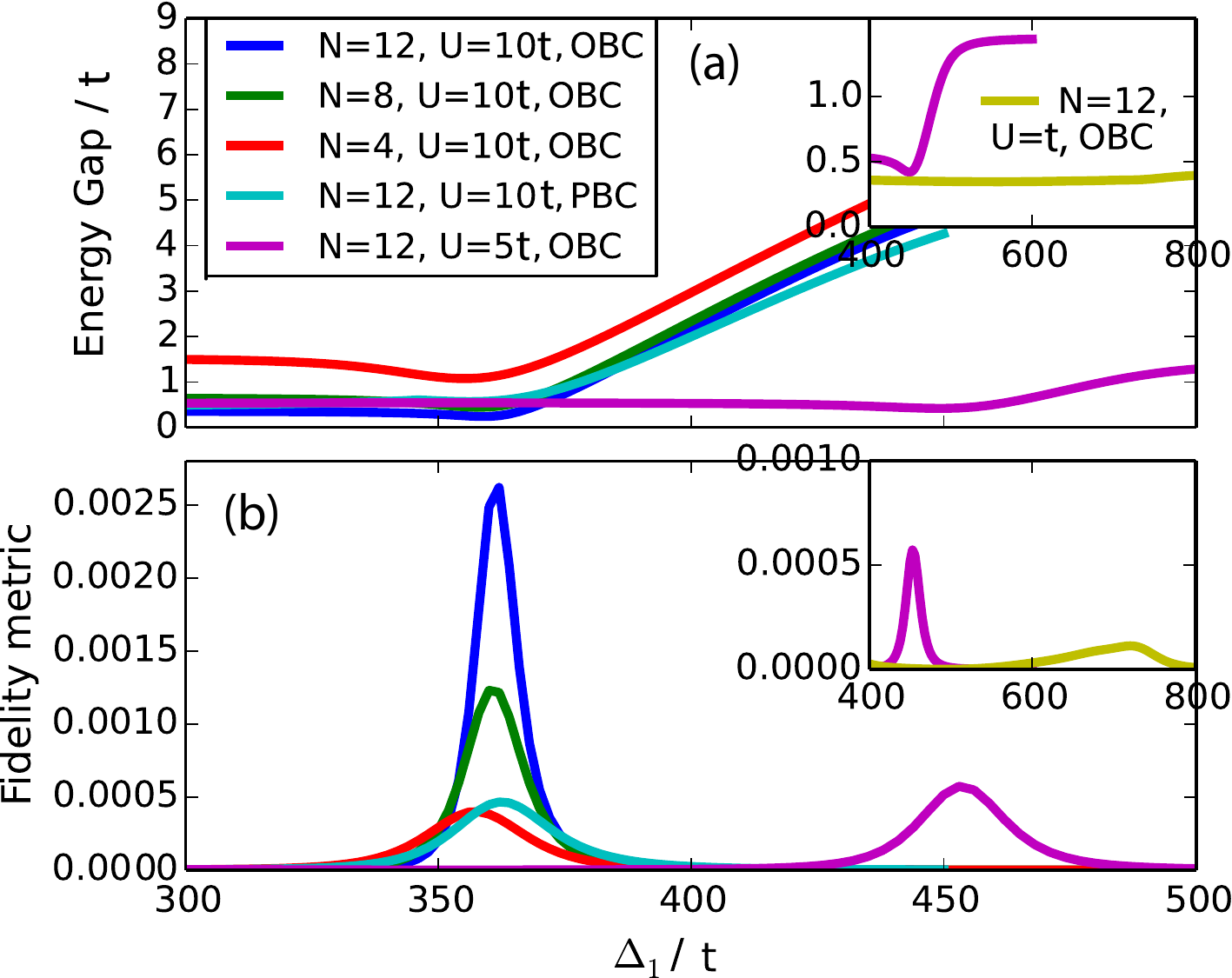}\caption{(a) Energy
gap for different values of $U$ and $N$. The inset shows a regime when $U=t$,
in yellow, for $N=12$ with OBC compared to $U=5t$ from the main figure. (b)
The peaks of Fidelity metric illustrate the critical points. When $N$ varies,
the location of the critical point remains intact. However, varying the
on-site interaction $U$ changes the location of the critical point, which is
consistent with the analysis in Sec. \ref{sec_numerical}.}%
\label{Incorporated}%
\end{figure}

\begin{figure}[th]
\centering
\includegraphics[width=\columnwidth]{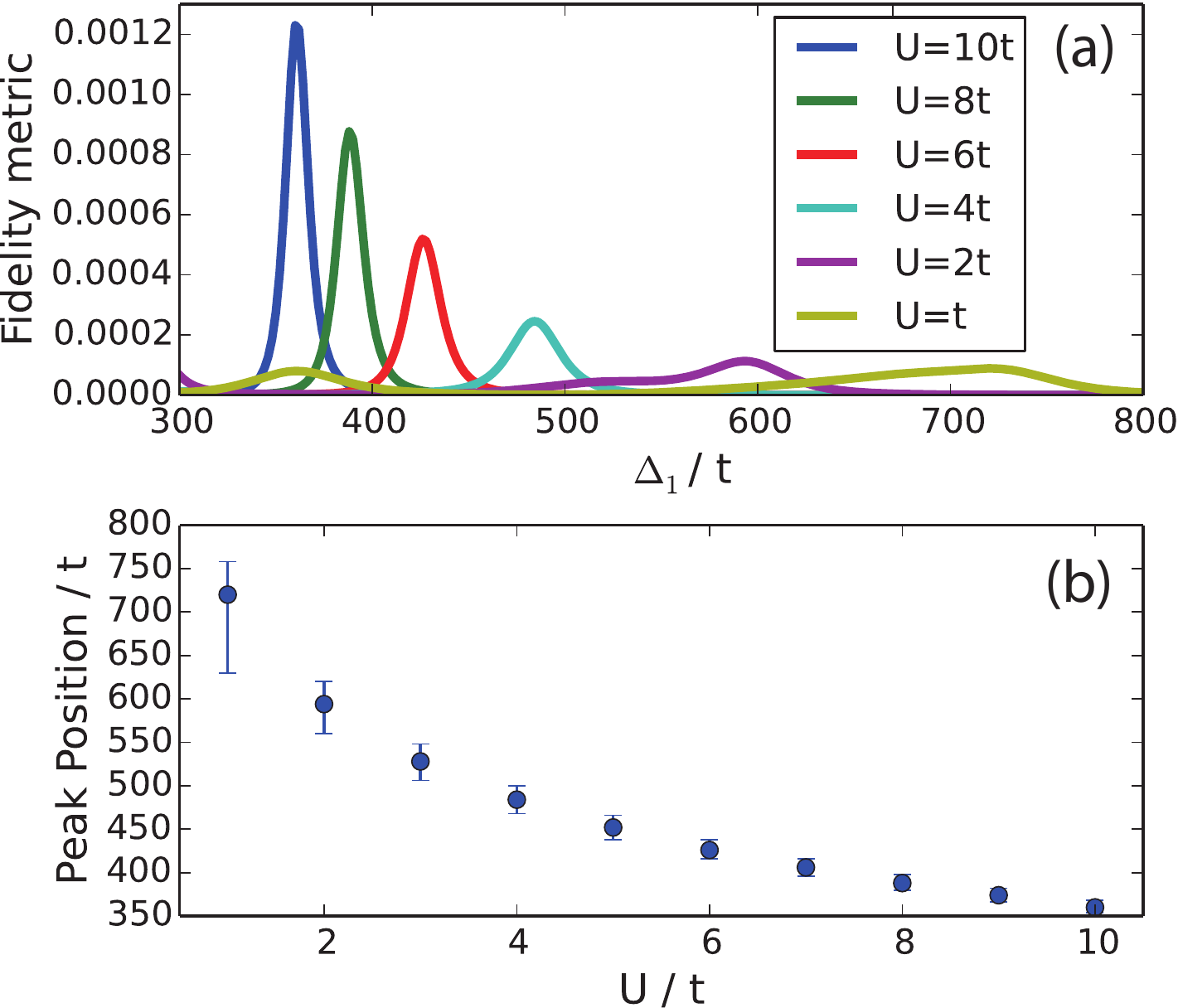}\caption{(a)
Fidelity metric as a function of $\Delta_{1}$ for different values of $U$ for
$N=8$ and $7$ photons. (b) Peak position of Fidelity metric as a function of
$U/t$. The full width at half maximum (FWHM) is shown as the bar spanning
across each point. }%
\label{FidelityMetric}%
\end{figure}

For a homogeneous 1D array of $N$ sites, the $(N-1)$ photons are not localized
if the hopping coefficient is finite. By increasing the on-site energy of the
first site, it becomes unfavorable if any particle hops into it. If the
repulsive interactions between the bosons exceed the critical value of the
MI-SF transition ($U_{c}/t\approx3.28$ in 1D \cite{1DBHM PRA, 1DBHM EPL}), the
ground state for the rest $N-1$ sites becomes a Mott insulator with a
wavefunction in Fock space as
\begin{equation}
\left\vert \varphi_{1}\right\rangle =\left\vert 0,1,1,...,1\right\rangle .
\end{equation}
By applying this ground state to the Hamiltonian (\ref{Site wise}), one gets
the ground state energy
\begin{equation}
E_{1}=\left\langle \varphi_{1}\right\vert H\left\vert \varphi_{1}\right\rangle
=-\mu(N-1).
\end{equation}

Then we estimate the ground state of a SF and compare the two ground state
energies to determine where the transition occurs when $\Delta_{1}$ is varied.
In our mean-field approximation, we consider a simplified trial ground state
with no double occupancy, which is appropriate for the case $U\gg t$. In Fock
space, states like $\left\vert 0,2,0,1,...,1\right\rangle $ are neglected.
Thus the trial ground state is
\begin{align}
|\varphi_{2}\rangle &  =\frac{1}{\sqrt{N}}(\left\vert 0,1,1,...,1\right\rangle
+\left\vert 1,0,1,...,1\right\rangle \nonumber\\
&  +\left\vert 1,1,0,...,1\right\rangle +...+\left\vert
1,1,1,...,0\right\rangle ).
\end{align}
The ground state energy is
\begin{align}
E_{2}  &  =\left\langle \varphi_{2}\right\vert H\left\vert \varphi
_{2}\right\rangle \nonumber\\
&  =\frac{1}{N}[-2t(N-1)-\mu N(N-1)+(\delta+\eta)(N-1)]\nonumber\\
&  \approx\delta+\eta-2t-\mu(N-1).
\end{align}
The energy difference between the two ground states is%
\begin{equation}
\Delta E=E_{1}-E_{2}\approx2t-(\delta+\eta). \label{critical_1}%
\end{equation}

A phase transition occurs at the crossing point $\Delta E=0,$ or $(\delta
+\eta)=2t$. Thus the system forms a Mott insulator by emptying the first site.
From Eq. (\ref{Varables}) we obtain an estimation of the phase transition
point at $\Delta_{1}\approx390t$ for $U=10t.$ To check this prediction and
provide more accurate estimations, we implement the ED method for several
moderate-size systems. Figures \ref{Ground state} and \ref{GroundstateCaseI}
show ground state properties including $n_{i}$ and $\sigma_{i}$ on different
sites as $\Delta_{1}$ varies. The energy gap of the first excited state, shown
in Figure \ref{GroundstateCaseI}(a), verifies the existence of the SF
(gapless) and MI (gapped) states.

The fidelity metric shown in Figures \ref{GroundstateCaseI}(b) and
\ref{Incorporated} captures and locates the critical regime when the on-site
energy of site 1 is manipulated. In Figure \ref{Ground state}, above
$\Delta_{1}/t\approx365$, the density is uniform away from site 1. The
variance $\sigma_{i}$ is also suppressed in the bulk. Thus the system is in
the MI regime. Below $\Delta_{1}/t\approx365$, the photons tend to congregate
at the two ends of the array, but the variance is small. At the center of the
array, the photon density is smaller with a larger variance. This corresponds
to a delocalized state. The density $n_{i}$ thus captures the main conclusion
of our mean-field analysis, and shows corrections from finite-size effects.

The critical values in the numerical results are close to the mean-field
estimations. The location of the critical point does not change much as $N$
changes, but the MI features become more prominent when $N$ increases. Due to
finite-size and boundary effects, the edge of the Mott insulator is distorted
but the bulk indeed exhibits features such as an integer filling and
suppressed fluctuations $\sigma_{i}$. Boundary effects can also be observed on
the neighbors of the manipulated site as their values of $n_{i}$ deviate from
the bulk. Those observations are also valid in Figure \ref{GroundstateCaseI}%
(a)(b), where site 1 is connected to site 2 and site 12 due to PBC.

For small $U/t$, as shown in Figure \ref{GroundstateCaseI} and the insets of
Figure \ref{Incorporated}, the SF state dominates the whole parameter space
explored in our ED calculations, which confirms that no artifact is induced if
the system is in the SF regime. In the insets of Figure \ref{Incorporated},
the results of a broader range of $\Delta_{1}$ for the case of $U=t$ is shown
and the small smooth gap through out the range of $\Delta_{1}$ is consistent
with a SF state of the case $U=t$ in Figure \ref{GroundstateCaseI}(e)(f).

Figure \ref{Incorporated} shows another signature of the phase transition as
$\Delta_{1}/t\approx365$ for $U=10t$ when $N=4,8,$ and $10$, as indicated by a
minimum in the energy gap followed by a rapid rise. For different values of
$U/t$, $\Delta_{i}$ in the bulk are different according to
Eq.~(\ref{Onsite Interaction}). Hence the critical point shifts in the
$\Delta_{1}/t$ axis according to Eqs. (\ref{Varables}) and (\ref{critical_1})
and this is consistent with the results shown in Figure \ref{Incorporated}.

\subsection{Case 2: $\rho=1$}

As illustrated in Figure \ref{Cases}(c)(d), here we consider $N$ photons
placed in an $N$-site array. If $U/t$ is large, the system is in a Mott
insulator state. As the on-site energy of site 1 increases, the boson in that
site is expected to be pushed to the bulk and this should lead to a
delocalized state because of the extra boson. Following a similar procedure,
we estimate the critical value of $\Delta_{1}$ that controls $\delta$ and
$\eta$ for this case.

\begin{figure}[th]
\centering
\includegraphics[width=\columnwidth]{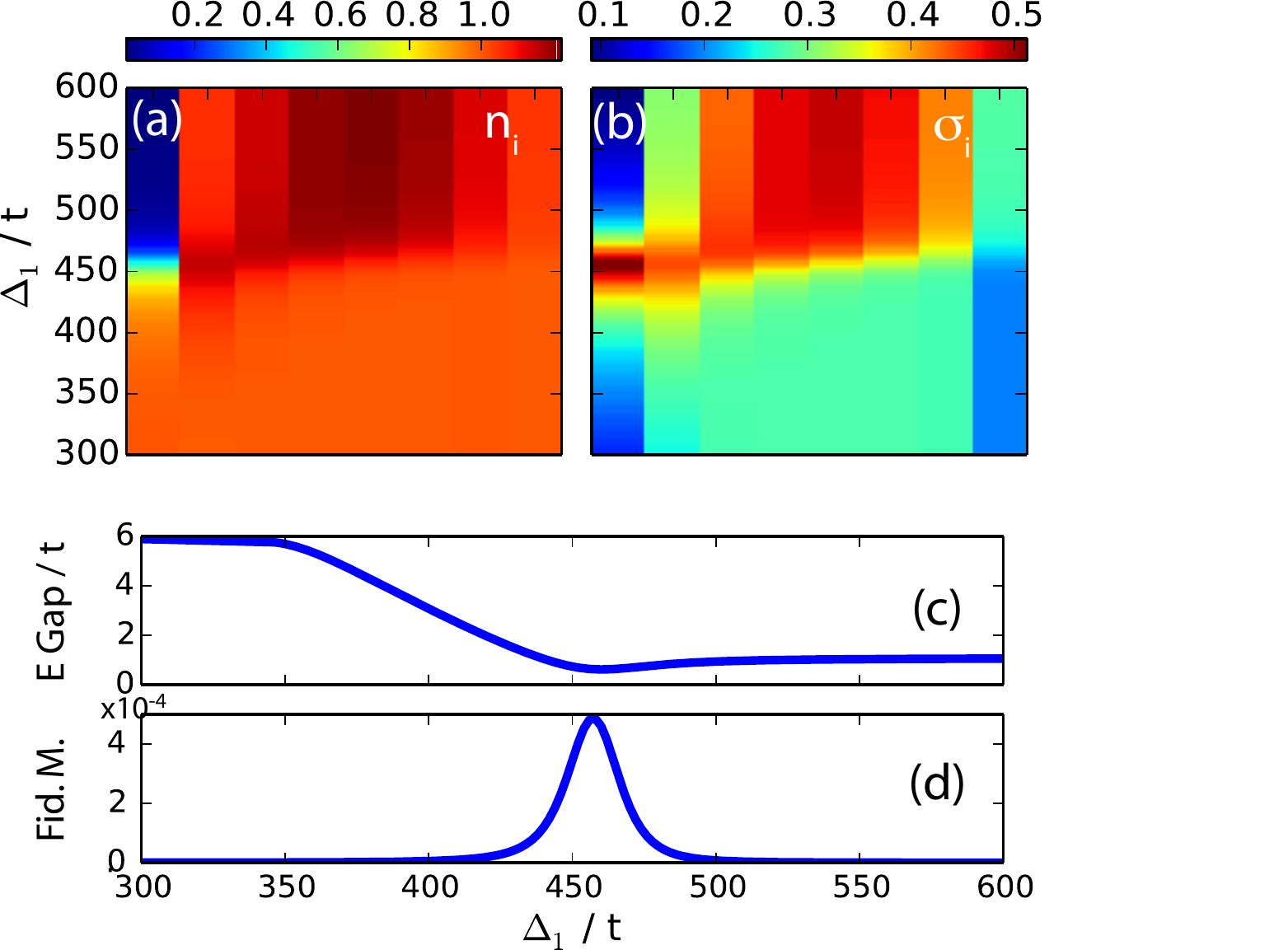} \caption{Exact
diagonalization results for Case 2 with $N=8$ and $8$ photons. Here $U=10t$.
(a) and (b) show the density profile in the array and the density variance.
The energy gap (E Gap) and fidelity metric (Fid. M.) in (c) and (d) clearly
exhibit signatures of the MI-SF transition. }%
\label{Case_2 diagram}%
\end{figure}

The localized MI ground state can be written as
\begin{equation}
\left\vert \varphi_{1}\right\rangle =\left\vert 1,1,1,...,1\right\rangle ,
\end{equation}
with the ground state energy
\[
E_{1}=\left\langle \varphi_{1}\right\vert H\left\vert \varphi_{1}\right\rangle
=\delta-N\mu.
\]
We consider a delocalized trial ground state
\begin{align}
\left\vert \varphi_{2}\right\rangle  &  =\frac{1}{\sqrt{N-1}}(\left\vert
0,2,1,...,1\right\rangle +\left\vert 0,1,2,...,1\right\rangle \nonumber\\
&  +...+\left\vert 0,1,1,...,2\right\rangle ),
\end{align}
whose ground state energy is
\begin{align}
E_{2}  &  =\left\langle \varphi_{2}\right\vert H\left\vert \varphi
_{2}\right\rangle \nonumber\\
&  =-\frac{(N-1)N}{N-1}\mu+\frac{N-1}{N-1}\frac{U}{2}-\frac{N-2}%
{N-1}2t\nonumber\\
&  \approx-N\mu+\frac{U}{2}-2t
\end{align}
Thus the energy difference is
\[
\Delta E=E_{1}-E_{2}\approx\delta-\frac{U}{2}+2t.
\]
The MI-SF phase transition occurs when $\Delta E=0$, and one may notice that
the critical point depends explicitly on $U$, which is in contrast to the
$U$-independent critical point in the mean-field analysis of case 1. For case
2 we obtain that the critical points are $\delta=3t,\Delta_{1}\approx469t$ for
$U/t=10$ and $\delta=0.5t,\Delta_{1}\approx470t$ for $U/t=5$.

Numerical results from the ED method for this case are shown in Figure
\ref{Case_2 diagram}. As shown in panels (a) and (b), below the critical point
$\Delta_{1}\sim470t$, the system is an MI with one photon per site and above
$\Delta_{1}\sim470t$ the system becomes an SF with significant $\sigma_{i}$ in
the bulk. The fidelity metric shown in panel (d) verifies that the critical
point is close to the estimation from our mean-field analysis. These results
verify the feasibility of inducing and observing those transitions in
moderate-sized systems.

\section{Implications for experimental realization}

\textit{State Preparation:} In the MI regime, the particle density on each
site is an integer. One may prepare an arbitrary $n$-photon state in each
site, including $n=0,1$ that are of interest, by adiabatically swapping the
qubit state to the TLR \cite{Single Photon,ArbitaryNstate TLR}. This single
site preparation can be performed simultaneously on all the sites. Then
starting from the MI regime, one can transform it to the many-body ground
state for different cases. For example, in case 1 in Sec.III, the ground state
in the MI regime is $\left\vert 0,1,1,1,...\right\rangle $. Recent work also
proposes a scheme of a $N$ photon state preparation in a superconducting TLR
array supported by numerical results \cite{JCHM SC chain}.

\textit{Cooling:} Solid state simulators based on superconducting circuits
including the one we propose here contain many degrees of freedom, which not
only provide great tunability but also introduce relatively strong couplings
to external fields. To experimentally implement the simulator proposed here,
cooling such a complex system can be a great challenge. We suggest the
following three stages. In stage 1, the whole system is kept in the
superconducting phase and thermal excitations in the superconducting circuits
and Josephson junctions should be suppressed
\cite{MakhlinReview,DevoretReview,DevoretReview2,YouReview,ClarkeReview}. They
are also associated with suppression of dissipation and decoherence. As
mentioned in the introduction, the life time of the photons at this stage is
already much longer than the operation time of the superconducting circuit by
a factor about $10^{7}$.

In stage 2, cooling of the TLR-qubit single site system should be performed
before connecting the whole array. This is associated with the state
preparation of the TLR array and a different degree of freedom from that of
stage 1 needs to be dealt with. The quantum computation community has been
making significant progresses related to the cooling at this stage
\cite{YouReview}. Inspired by ideas from optical systems, Sisyphus cooling and
side-band cooling of superconducting systems have successfully cooled a qubit
to its ground state \cite{Sisyphus Cooling, Sideband Cooling, You Cooling}.

In stage 3, once a multi-site array is connected by turning on the hopping
between adjacent sites, the desired many-body Hamiltonian follows. In order to
simulate and observe the quantum phase transition discussed here, one needs to
constantly cool the system and keep the number of photons conserved during the
operation. This is more challenging than cooling just a single site,
especially inhomogeneity of the on-site energies is present. Applying a bias
or other manipulations of the parameters can cause excitations as well and
need to be performed with care. Moreover, to take out the heat from the
multi-site system when operating near the critical regime leads to yet another
issue. Advanced schemes for cooling a single site have been available while
cooling a multi-site array like the one studied here has not been reported so
far. Development of such technologies is important for realizing the proposed
simulator. Based on current ground-state preparations and state-manipulation
technologies developed in coupled superconducting cavity systems \cite{Matteo
Shellgame, Chen Simulation}, it is promising that photon-number-conserving
ground-state cooling processes may be realized by scaling up the cooling
methods for those coupled systems.

\begin{figure}[th]
\centering
\includegraphics[width = 0.7\columnwidth]{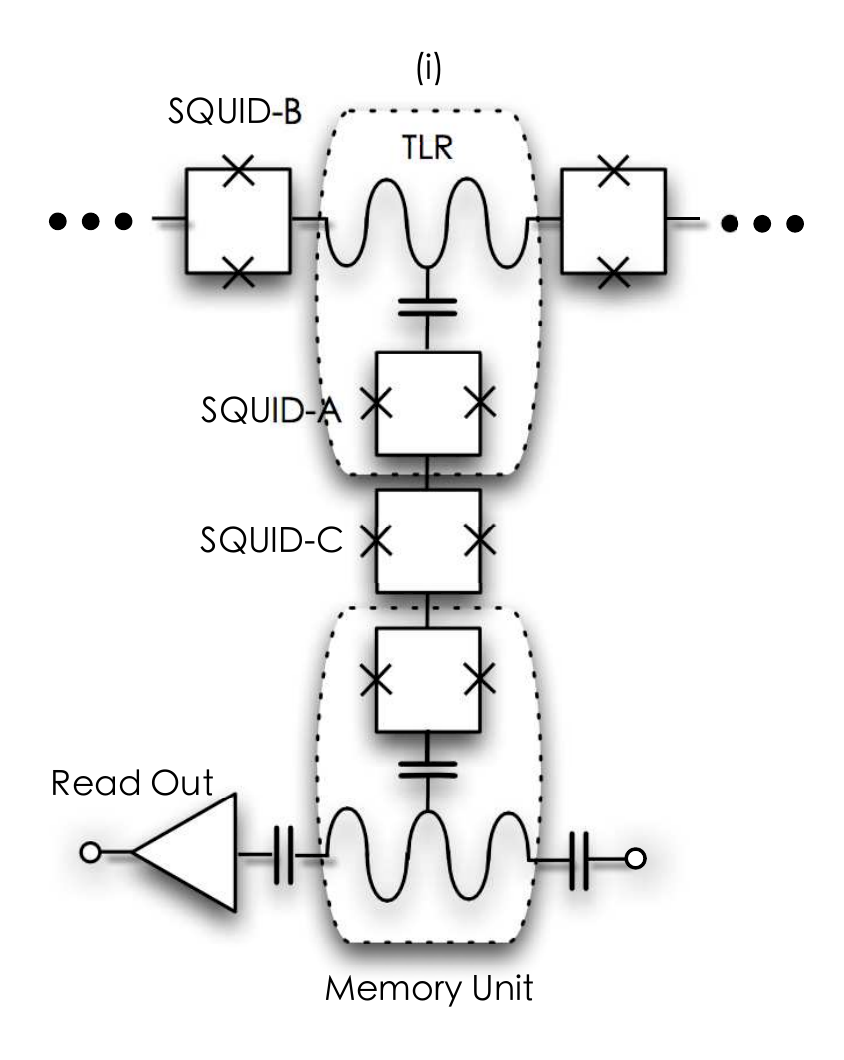}\caption{Measuring
the photons in the simulator: Each site of the simulator is connected to a
memory unit formed by another qubit-TLR system via a tunable SQUID (labeled as
SQUID C) acting as a switch. Measurements of the photon number in the memory
unit can be applied\cite{Detect Lang,Detect Riste, Yin RFphoton}. This memory
unit can also serve as a circuit for preparing the initial state by
manipulating SQUID-C and SQUID-B. }%
\label{Measure Circuit}%
\end{figure}

\textit{Detection of phase transition: } Since the single-site manipulations
of the MI-SF transition exhibit strong signatures in the density distribution,
we discuss a direct measurement of the photon numbers and number fluctuations
on each site. Interestingly, the measurement can be turned on and off when
needed. As shown in Figure \ref{Measure Circuit}, each site can be coupled to
a memory TLR via the additional circuit. The central SQUID-C is used to switch
the coupling between the on-site unit and the measurement unit \cite{Tunable
Coupler Science} for controlling the memorizing window. This is possible by
changing the bias flux through SQUID-C (labeled on Figure
\ref{Measure Circuit}), $\phi_{m}$. A fast photon state SWAP between the two
TLRs can be applied with four-wave mixing \cite{FWMtoolbox} to get $\left\vert
n_{on-site}0_{measure}\right\rangle \rightarrow$ $\left\vert 0_{on-site}%
n_{measure}\right\rangle $, so that the photons in the TLR of the simulator
are transferred and stored into the measurement TLR. Single photon state fast
measurements can be applied to measure photon numbers in the memory TLR with
technologies developed in circuit QED recently \cite{Detect Lang,Detect Riste,
BozyigitWallraffNPhys2010,JohnsonNPhys2010,CDengLupascu, Yin RFphoton}. By
repeating the measurement one gets the average photon number $\left\langle
n_{i}\right\rangle $ and variation $\left\langle \sigma_{i}\right\rangle $ as
depicted in Figure \ref{Ground state} for detecting different quantum phases
in the TLR array.

To summarize, a promising way to realize this simulation is: (1) Tune the
parameters in the MI regime and prepare the array in the ground state with an
integer number of photons. (2) Adiabatically adjust the parameters to the
desired values and cool the photons down to their ground state within the
photon relaxation time. (3) Measure the photon number in each single site.
Then repeat (1) to (3) to obtain the average photon number and number fluctuations.

\section{Conclusion}

A versatile quantum simulator of interacting bosons based on a tunable
superconducting TLR-SQUID array has been presented. The BHM with tunable
parameters on each site can be studied using the photons in this simulator. We
have demonstrated the feasibility of inducing the MI-SF transition by
manipulating only one single site. Our results are further supported by the
exact diagnolization method, and details of the transition with realistic
parameters are presented. The fidelity metric, energy gap, and on-site photon
number show signatures of the phase transition. We also discussed possible
schemes for state preparation, cooling, and detection of the phase transition
for this proposed simulator.

Besides the manipulations of the phase transition discussed here, this quantum
simulator is also capable of demonstrating topological properties in the BHM
with superlattice structures and should exhibit the topological properties,
edge states, and topological phase transitions studied in
Refs.~\onlinecite{SLiang1D,Edgestate 1D
Superlattice, Zakphase}. Moreover, quantum quenches \cite{Quench, One
dimensional boson} and their associated dynamics may also be simulated by this
superconducting circuit simulator as well. For example, similar to
Ref.~\onlinecite{Quenchcurrent} one can separate the TLR array into two
sections by turning off the hopping between the two sections. Then different
photon numbers are prepared in the two sections. By switching on the hopping
between the two sections, photons are expected to slosh back and forth between
the two sections, which should be detectable with similar measurement methods.
Thus the superconducting circuit simulator adds more excitement to the physics
of interacting bosons and complements other available simulators.

\section{Acknowledgments}

We thank Lin Tian and Raymond Chiao for valuable advice on this work and Kevin
Mitchell and Jay Sharping for useful discussions. The computations were
performed using the resources of the National Energy Research Scientific
Computing Center (NERSC) supported by the U.S. Department of Energy, Office of
Science, under Contract No. DE-AC02-05CH11231.

\end{document}